\PassOptionsToPackage{desactivate}{linenoaa}
\documentclass{aa}
\usepackage{graphicx} 
\usepackage{hyperref}
\usepackage[dvipsnames]{xcolor}
\usepackage{float}

\newcommand{\byol}{\texttt{BYOL}~}

\renewcommand{\texttt}[1]{%
  \begingroup
  \ttfamily
  \begingroup\lccode`~=`/\lowercase{\endgroup\def~}{/\discretionary{}{}{}}%
  \begingroup\lccode`~=`[\lowercase{\endgroup\def~}{[\discretionary{}{}{}}%
  \begingroup\lccode`~=`.\lowercase{\endgroup\def~}{.\discretionary{}{}{}}%
  \begingroup\lccode`~=`_\lowercase{\endgroup\def~}{_\discretionary{}{}{}}%
  \catcode`/=\active\catcode`[=\active\catcode`.=\active\catcode`_=\active
  \scantokens{#1\noexpand}%
  \endgroup
}

\begin{document}

\title{\texttt{astromorph}: self-supervised machine learning pipeline for astronomical morphology analysis}

\author{
    P. Bjerkeli\inst{\ref{inst1}}$^{\dagger}$
    \and J. Kainulainen\inst{\ref{inst1}}$^{\dagger}$ 
    \and M.~C. Toribio\inst{\ref{inst2}}$^{\dagger}$ 
    \and L. Boschman\inst{\ref{inst3}}
    \and O. Maya Lucas\inst{\ref{inst3}}
}

\institute{
    Department of Physics and Astronomy, Chalmers University of Technology, 412 96 Gothenburg, Sweden\label{inst1}
    \and Department of Physics and Astronomy, Chalmers University of Technology, Onsala Space Observatory, 439 92 Onsala, Sweden\label{inst2}
    \and Chalmers e-Commons, Chalmers University of Technology, 412 96 Gothenburg, Sweden\label{inst3}
    \\
    \\ \email{per.bjerkeli@chalmers.se}
    \\ \\
    ${\dagger}$ Authors contributed equally to this work.
}

\date{Submitted}
 \date{Submitted October 10, 2025; accepted February 6, 2026}

\abstract
{Modern telescopes generate increasingly large and diverse datasets, often consisting of complex and morphologically rich structures. To efficiently explore such data requires automated methods that can extract and organize physically meaningful information, ideally without the need for extensive manual interaction.}
{We aim to provide a user-friendly implementation of a self-supervised machine learning framework to explore morphological properties of large datasets, based on the BYOL (Bootstrap Your Own Latents) method. By enabling the generation of meaningful image embeddings without manually labelled data, the framework will enable key tasks such as clustering, anomaly detection, and similarity based exploration. }
{We present \texttt{astromorph}, a Python package that implements the BYOL method in a way tailored for astronomical imaging. In contrast to existing BYOL implementations, \texttt{astromorph} accommodates data of varying dimensions and resolutions, including both single-channel FITS images and multi-channel spectral cubes. The package is built with usability in mind, offering streamlined pipeline scripts for ease of use as well as deeper customization options via PyTorch-based classes. }
{To demonstrate the utility of \texttt{astromorph}, we apply it in two contrasting science cases representing different astronomical domains: images of protoplanetary disks observed with the Atacama Large Millimeter/submillimeter Array (ALMA), and infrared dark clouds observed with Spitzer and Herschel. In both cases, we demonstrate how \texttt{astromorph} produces scientifically meaningful embeddings that capture morphological differences and similarities across large samples. 

}
{\texttt{astromorph} enables users to apply a robust, label-free approach for uncovering morphological patterns in astronomical datasets. The successful application to two markedly different datasets suggest that the pipeline is broadly applicable across a wide range of imaging-rich astronomical context, providing a user friendly tool for advancing discovery in observational astronomy. }


\authorrunning{Bjerkeli, Kainulainen, Toribio, Boschman, and Maya Lucas}
\titlerunning{\texttt{astromorph}}

\maketitle

\section{Introduction}
\label{sec:introduction}
Astronomical data sets are commonly large, complex, and difficult to interpret in terms of physically meaningful characteristics. These vast amounts of data are routinely gathered and stored in publicly available archives. Efficient use of such data requires approaches and tools to automatically explore and identify potentially interesting information. In recent years, especially machine learning based approaches have emerged to harvest science from the large data sets and archives.

One phenomenon common for many topics in astronomy is that the physical properties of an object can have a crucial impact on object’s geometric appearance, i.e., on its morphology. Consequently, observations and analysis of object morphology is a standard approach for a wide spectrum of science questions. Machine learning approaches that exploit convolutional neural networks (CNNs) are inherently well-suited to address such questions. In general, the CNNs work by converting an image into a feature vector (aka embedding) that represents the properties of the image in an abstract manner. These vectors, or embeddings, are then the basis for further analyses, e.g., for categorization.

A common problem that hampers CNN-based approaches in astronomy is that the available training data is usually limited, and that it is not desirable to impose labelling on the data beforehand. Unsupervised approaches, such as self-organizing maps (SOMs), have recently been applied to cluster or characterize astronomical data without labels \citep[e.g.,][]{Vantyghem2024}. A complementary strategy is self-supervised learning, which likewise avoids the need for labelled training data while enabling the training of deep neural networks on large datasets. In particular, The Bootstrap Your Own Latents (BYOL; Grill et al. 2020) method is a self-supervised representation learning method that does not require labelling of the training data, thus removing a manual process from the analysis. Its effectiveness in producing useful embeddings in an astrophysical context has been demonstrated by Mohale \& Lochner (2024), who used BYOL to create embeddings that are useful for classification of galaxies.

Clearly, the approach of using BYOL can be thought to be applicable to a wide range of science questions that regard the morphology of an object. However, a limiting factor for widespread application of BYOL in astronomy is that the existing BYOL implementations are tailored towards (pre-trained) models for terrestrial images. The implementations assume that all the images in the data set will be 3-channel RGB images of the same size; in astronomy, images can have all possible shapes and sizes. Similarly, astronomical images have a wide variety in their number of channels, ranging from the conventional single-channel images to the data cubes common in radio astronomy. Finally, few astronomers are experts in machine learning and adapting the approach to the needs of astronomical  context can be a formidable task.

To address the above issues, we present and release to the community a package that implements the BYOL method, called \texttt{astromorph}. Our package aims to make BYOL easily accessible to any astronomer by removing some of the more tedious parts of machine learning and providing a framework prepared particularly for astronomical context. 
The \texttt{astromorph} package is available on GitHub (\href{https://github.com/onsala-space-observatory/astromorph}{github.com/onsala-space-observatory/astromorph}) and distributed via PyPI (\href{https://pypi.org/project/astromorph/}{pypi.org/project/astromorph/}). Installation is performed using \texttt{pip install astromorph}.

We will briefly demonstrate the utility and functionality of \texttt{astromorph} in the context of two inherently different science cases. The first regards the morphology of protoplanetary disks as observed by the Atacama Large Millimeter/submillimeter Array (ALMA). During the last decade, ALMA has observed thousands of protoplanetary disks. One important question regards the differentiation in observed disk morphologies, linked to physical processes in the disk such as ring formation due to planet formation or dust concentrations due to gravitational instabilities. We will use \texttt{astromorph} to automatically classify disks from the ALMA archive and to search for disks that have morphologies similar to objects of particular interest.

The second science question regards the morphology of dense star-forming molecular clouds in the Milky Way. We currently believe that processes such as gravity, turbulence, and magnetic fields leave their fingerprints on cloud morphology during the evolutionary life-cycle of the clouds. One avenue to constrain these processes is to identify those fingerprints from the cloud morphology, which can then be used to constrain the evolutionary processes. We will use \texttt{astromorph} to analyse the morphology of thousands of molecular clouds and demonstrate that the resulting embeddings enable the investigation of potentially physically meaningful morphological patterns.

With the above two science cases, we demonstrate the use of \texttt{astromorph}, and hence the BYOL method, in finding objects-of-interest from a large archive (disks) and in finding patterns from data in an unsupervised manner (clouds). Similar questions repeat themselves across astronomical topics; one can envision \texttt{astromorph} being useful for a plethora of applications in which the morphology of objects is of scientific interest.

The paper is organised as follows. In Section \ref{sec:methods} we describe how we build the \texttt{astromorph} package to implement the BYOL method. In Section \ref{sec:hyperparameters} we explore and discuss the various parameter/method choices done during the implementation. Section \ref{sec:results} presents our two science demonstration cases. Section \ref{sec:conclusions} summarises and concludes our work.  

\section{Methods}
\label{sec:methods}
In this paper, we develop and present a software package that implements the BYOL method \citep{grill2020}. Our package is heavily inspired by the PyTorch-based \texttt{byol-pytorch} package\footnote{\url{https://github.com/lucidrains/byol-pytorch}}, but to increase flexibility and to better accommodate astronomical data with a number of channels differing from the typical terrestrial 3-channel RGB images,
we write our own package to implement the BYOL\footnote{For clarity, we use roman type when referring to BYOL as a methodological approach, and we use typewriter capitals when referring to code or script names (e.g. the \texttt{BYOL} class in \texttt{astromorph}).}. 
The goal of our implementation is to provide a user-friendly package that is as such suitable for astronomical context and that abstracts away, or helps the user to  apply, some of the more tedious or complex parts of the machine learning process. Therefore, the \texttt{astromorph} package is structured to support common astronomical data analysis tasks through a streamlined pipeline, while still allowing expert users to customize the workflow as needed.

Overall, \texttt{astromorph} can be regarded as a set of wrappers that enclose the basic BYOL architecture and provides the user simplified ways to apply it. Figure \ref{fig:astromorph-pipeline} shows a process diagram describing \texttt{astromorph} at a high level. We use the diagram in the following sections as the basis for explaining how \texttt{astromorph} implements the BYOL method and how the user can choose different models of operation depending on their desired level of interaction or expertise. Later, Section \ref{sec:hyperparameters} presents and discusses in detail the parameters and user choices available in our implementation.

\subsection{The BYOL method}
\label{sec:byol}

\begin{figure*}
    \centering
     \includegraphics[width=1.05\linewidth]{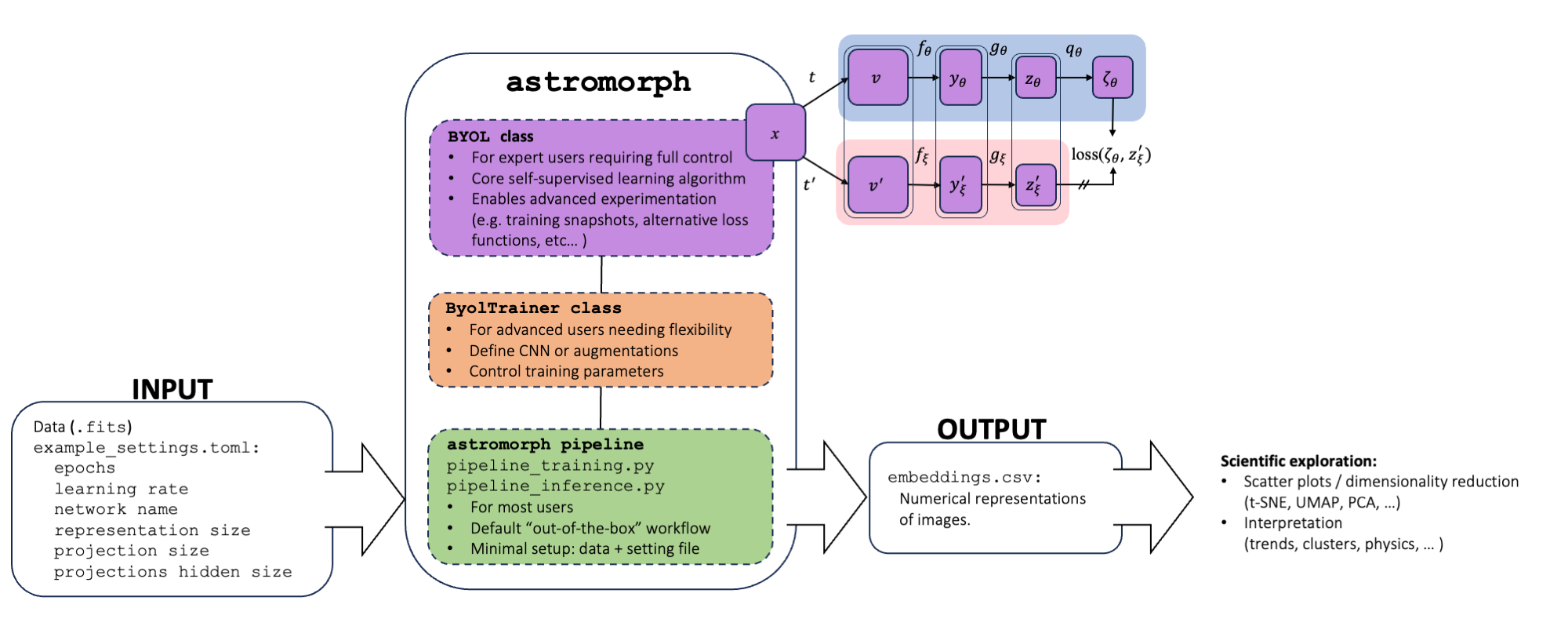}
    \caption{Process diagram of the \texttt{astromorph} package.
    The \texttt{pipeline\_training.py} script builds and trains a CNN model using the BYOL framework, with configuration defined in \texttt{example\char`_settings.toml}. The BYOL architecture (upper right) operates on two augmented views on the input to learn meaningful representations without labels. The script \texttt{pipeline\_inference.py} uses the trained model to compute and store embeddings that can be used for further scientific exploration.}
    \label{fig:astromorph-pipeline}
\end{figure*}

The BYOL method is described in \citet{grill2020}, to which we refer for a full technical description. We summarise and discuss the method here to the degree necessary to provide context for the description of the \texttt{astromorph} package.

The BYOL method is a self-supervised representation learning approach, related to contrastive learning but distinct in that it does not require negative pairs. In contrastive learning, the neural network is presented with pairs of images. Most methods employ training pairs that can be considered either positive, in which case the images are related, or negative, in which case they are not. The BYOL method differs from this in that it achieves learning using only positive pairs; 
these pairs are generated from the original images via stochastic image transformations, leading to two views of each image.
Examples of such transformations include a horizontal flip applied with a probability of 0.5, or a random rotation sampled from 0 to 360 degrees. Several of these stochastic transformations are applied in sequence, so that repeated application of the transformation sequence leads to two different views of the same object.
An important note is that these transformations should maintain the essential properties of the object in the image that will be used for its morphological analysis. 
For example, if orientation with respect to some external axis is an important quality of the object in the image, rotation of the image will change that property, and should therefore be excluded from the transformation sequence. Thus, especially the more expert use of the method will have to carefully consider the role of the transformations.

Figure \ref{fig:astromorph-pipeline} illustrates the overall architecture of our software. The figure includes  
a schematic overview of the BYOL framework adopted and simplified from \citet{grill2020}.
It consists of two components: an online component (light blue) and the target component (pink). The online component consists of a CNN ($f_\theta$), a projector ($g_\theta$), and a predictor ($q_\theta$), while the target component contains the (same) CNN ($f_\xi$) and projector ($g_\xi$). The parameters of the target component are updated to follow the parameters of the online component using an exponentially weighted moving average, which means that the target changes slowly and lags behind the online component. Training is achieved by comparing and optimizing the output from the predictor of the online component and the output from the projector of the target component. In this optimization, we use the negative cosine similarity as the loss function. 

For a user to apply the BYOL method in a scientifically meaningful way, numerous auxiliary operations, methodological choices, and parameter settings are required. For example, these include preparation and input of the scientific data, setting up and scripting the learning process, and deciding on  methodological details such as data augmentation strategies and training parameters. To streamline these steps and adapt them to an astronomical context, we encapsulated the method into python scripts and classes that together form the \texttt{astromorph} package.  

\subsection{The \texttt{\normalfont \texttt{astromorph}} package} 
\label{sec:astromorph}

The \texttt{astromorph} package is a set of python classes and scripts that provide abstractions for implementing the BYOL method. At its core lies the \texttt{BYOL} class, which implements the BYOL method; other classes and scripts are wrappers that ultimately call this \texttt{BYOL} class (see Fig. \ref{fig:astromorph-pipeline}). These supporting classes and scripts also contain default parameter and input settings, chosen to offer a reasonable starting point for a wide range of problems in an astronomical context (see Section \ref{sec:hyperparameters} for a detailed introduction/discussion of these choices). 

In the following, we describe the functionality of \texttt{astromorph} beginning with the simplest way to run it; we call this the \texttt{astromorph} pipeline. We then provide the details for users to better interact with the training process. This is done by specifying additional options while using the \texttt{ByolTrainer} class. Next, we describe the most advanced option to adapt the method by directly making use of the BYOL class itself, which provides numerous configuration possibilities for expert users. The relation between these levels of use of our package is illustrated in Fig. \ref{fig:astromorph-pipeline}. Finally, we introduce our light-weight CNN AstroMorphologyModel, developed to be used conveniently in this framework.

\subsubsection{The \texttt{astromorph} pipeline}
\label{sec:astromorph_pipeline}
The \texttt{astromorph} pipeline provides the simplest way to run our BYOL implementation. The pipeline is run through two Python scripts that can be invoked from the command line. These scripts are \texttt{pipeline\char`_01\char`_training.py} and \texttt{pipeline\char`_02\char`_inference.py}. The former executes the training of the model, and once done, embeddings can be obtained by running the latter. The embeddings can then be visualised in TensorBoard \citep{tensorflow2015}, or exported to a CSV-file for further analysis.

Basic customization of the pipeline -- for example setting the number of epochs or the learning rate -- is possible through a configuration file \texttt{example\char`_settings.toml}. To execute the pipeline the user only needs to modify that parameter file according to their needs. However, for any more refined customization than the basics, we recommend using either the \texttt{ByolTrainer} or \texttt{BYOL} classes as described in the next two sections \ref{sec:byoltrainer} and \ref{sec:byol-class}.

\subsubsection{The \texttt{ByolTrainer} class}
\label{sec:byoltrainer}
The training script of the \texttt{astromorph} pipeline uses several fixed methodological and technical choices that more advanced users may wish modify. To enable such users to have more control over the training process, we provide the class \texttt{ByolTrainer}.  

To make use of the \texttt{ByolTrainer} class, at a minimum, the user needs to set up their own scripts to provide the input data and provide a \texttt{pytorch}-based CNN. The user must specify which layer of the network produces the embeddings using the texttt{hidden\_layer} parameter (default: "\texttt{avgpool}"), and define the dimensionality of the embedding vectors via the \texttt{representation\_size} parameter (default: 128). Other options are possible, but then more expert modification of the \texttt{ByolTrainer} is needed; also, if the user feels that significant changes to the \texttt{ByolTrainer} are needed, it may be best to not use the \texttt{ByolTrainer} at all, but to write one's own learning script that directly makes use of the \texttt{BYOL} class  (see \S\ref{sec:byol-class} below).

The \texttt{ByolTrainer} class provides a simple default augmentation; if none is specified, users should check whether this default is suitable and modify if needed. 

We foresee that the typical use case for the \texttt{ByolTrainer} class (instead of the \texttt{astromorph} pipeline) is when one wishes to use a CNN or data augmentation different from the default ones (RandomHorizontalFlip, RandomRotation, randomly applied GaussianBlur). This could be the case when the orientation of the object is important, as explained in Section \ref{sec:byol}. Another situation can be if the user wishes to have more control on how the training and test data sets are treated; in this case the user would write their own scripts for data treatment and call the \texttt{ByolTrainer} class from them.

\subsubsection{The \texttt{BYOL} class}
\label{sec:byol-class}
The \texttt{astromorph} package implements the BYOL method \citep{grill2020} in a python class named \texttt{BYOL}. We have written our own implementation of BYOL because the existing implementation by \cite{grill2020} has limitations when applied to astronomical data, in particular, it is limited to two-dimensional, three-channel images; \texttt{astromorph} does not have this limitation. 

Our \texttt{BYOL} class wraps around a user-provided CNN and provides it with the functionality necessary for training the network through the BYOL framework.
It can return either the error value ($\mathrm{loss} ( \zeta_{\theta}, z'_{\xi} )$ in Fig. \ref{fig:astromorph-pipeline}) during training or the embeddings ($y_{\theta}$) during inference.

The \texttt{BYOL} class forms the core of the \texttt{astromorph} package and is used when running either the \texttt{astromorph} pipeline (\S\ref{sec:astromorph_pipeline}) or the \texttt{ByolTrainer} class (\S\ref{sec:byoltrainer}). A user with expertise in machine learning may also use the \texttt{BYOL} class without the \texttt{astromorph} pipeline or the \texttt{ByolTrainer} class; doing so requires a good level of expertise in machine learning, the BYOL method, and the PyTorch library. Typical examples in which an expert user might want to do this include cases where they want tight control over the training process -- for example, saving checkpoints during training, experimenting with alternative loss calculation methods, or applying BYOL to 3D spectral data. Below, we explain our implementation of the \texttt{BYOL} class and how to adjust it. 

The \texttt{BYOL} class is a subclass of the PyTorch \texttt{Module} class, and can therefore be interacted with in the same way as other PyTorch-based neural networks. The class expects that the base network (usually, the CNN), will be provided as a subclass of the \texttt{torch.nn.Module} baseclass. The user can provide their own neural network, or make use of a pre-defined one.

The existing \texttt{byol-pytorch} implementation expects all images within a dataset to have the same dimensions. This can be justified by making batch normalization easier, but the \texttt{byol-pytorch} implementation enforces this restriction even for models without batch normalization, where it is not actually required. One adjustment in our \texttt{BYOL} class compared to the existing ones is how input images are handled. Batch normalization is commonly used in CNNs trained with BYOL, but it is not inherent to the method itself; a network can be trained with BYOL without batch normalization, as long as the images can still be stacked into batches. However, with astronomical images one often needs to work with differing sizes and aspect ratios, even within a single dataset. To overcome the problem of stacking images with different aspect ratios, we use a combination of mirroring and rotation around the z-axis to create four different copies of the same image, all with the same aspect ratio.
We implement this in the \texttt{FilelistDataset} class where the user only needs to provide a list of the original FITS filenames. This class also provides the ability to copy a single-channel image into a 3-channel image for transfer learning of an RGB-oriented pre-trained network.

The \texttt{BYOL} class can be run with the default options by specifying as input only the CNN that will be used in conjunction with the BYOL method and the size of the resulting embedding vectors. It can also be adjusted via options that can be specified via keywords, but that are not strictly necessary for the code to run. These options are: 
\begin{itemize}
    \item the layer of the neural network to be intercepted;
    \item the stochastic augmentation function;
    \item the sizes of the hidden and output layer of the projector and predictor;
    \item the loss function;
    \item the decay parameter of the moving average function.
\end{itemize}
All but the first of these options are hyperparameters and are described in Sec.~\ref{sec:hyperparameters}. The interception layer is an architectural choice. In the case of transfer learning on top of a pre-trained neural network (e.g., the ResNet CNN), it is usually beneficial to specify a layer of the neural network to intercept. This interception is necessary, because typically the last layer of a pre-trained network is a prediction over $n$ categories. However, in our context we are not interested in the prediction into these categories, but rather in the underlying embedding vectors. To access these embeddings, we intercept the network at a stage before the layers responsible for making predictions.

\subsubsection{\texttt{AstroMorphologyModel} CNN}
Most pre-trained networks are computationally heavy, which can make transfer learning challenging on laptops and other non-dedicated devices. 
To accommodate users who have no experience building their own CNNs, we provide a light-weight CNN built using the \texttt{torchvision} framework \citep{torchvision2016}.
This network consists of several convolutional layers interspersed with batch normalization, ReLU, and a pooling layer, resulting in an embedding vector of size 128. This network achieves reasonable results for several types of data. It can also be trained on modest hardware, lowering the barrier for users who do not wish to start from large pre-trained models. However, for optimal results we still advise to fine-tune the model parameters, or use a pre-trained model.

\section{Hyperparameters}
\label{sec:hyperparameters}
Training a BYOL model requires several hyperparameters. Some directly influence the quality of the learned representations, while others are mainly technical but still necessary for the framework to run. We therefore divide the discussion into two groups: hyperparameters that affect model performance and those that are required for operation. In practice, basic hyperparameters can be set through configuration files used by the \texttt{astromorph} pipeline, more detailed options are exposed in the \texttt{ByolTrainer} class, and full flexibility is available in the \texttt{BYOL} class itself (see Table~\ref{tab:table1}).
\begin{table*}
    \centering
    \begin{tabular}{lccc}
    \hline\hline
       & \multicolumn{3}{c}{Input locations} \\
       \cline{2-4}
        Hyperparameter & \texttt{settings.toml} & \texttt{BYOLTrainer} & \texttt{BYOL} \\
       \hline
       Augmentation function  & N/A & \texttt{augmentation\_fuction} & \texttt{augmentation\_function} \\
       Learning rate   & N/A & \texttt{learning\_rate} & user defined \\
       Learning rate scheduler   & optional exponential decay & user defined & user defined \\
       Batch size & \texttt{batch\_size} & \texttt{batch\_size} & user defined \\
       Epochs & \texttt{epochs} & \texttt{epochs} & user defined \\
       Projection size & \texttt{projection\_size} & \texttt{projection\_size} & \texttt{projection\_size} \\
       Projection hidden size & \texttt{projection\_hidden\_size} & \texttt{projection\_hidden\_size} & \texttt{prediction\_hidden\_size} \\
       EWMA decay & N/A & N/A & \texttt{moving\_average\_decay} \\
       Loss function & N/A & N/A & \texttt{loss\_fn}\\
    \hline
    \end{tabular}
    \caption{Overview of key user-defined hyperparameters and where they can be specified when using the \texttt{astromorph} pipeline.}
    \label{tab:table1}
\end{table*}
\subsection{Influencing result}
\subsubsection{Augmentation function}
The augmentation function, denoted by $t$ in Fig. \ref{fig:astromorph-pipeline}, is a collection of stochastic transformations that are applied consecutively to create different views of the same image.
In the same figure, $t'$ is just a second application of the same augmentation function, which due to its stochastic nature will yield a different view every time. These two views are then used as a positive pair, with the goal that the neural network produces similar embeddings for both. Typical transformations include affine operations such as mirroring, arbitrary rotation, or shear.
The nature of these transformations should be such that an image before and after the transformation should still represent the same type of object.
For example, an upside-down transformation has serious implications for the meaning of terrestrial images, but this orientation does not matter so much in space. Similarly, rotations of astronomical images may or may not be important dependent on whether there is an external axis of relevance.

\subsubsection{Neural network}
The neural network ($f_{\theta}$ in Fig.~\ref{fig:astromorph-pipeline}) is the core of the entire framework, and is what we will eventually use for inference.
This package provides two types of neural networks, an \texttt{NLayerResnet} and \texttt{AstroMorphologyModel}.
\texttt{NLayerResnet} is a truncated ResNet18 \citep{2016cvpr.confE...1H}. This truncation is motivated by the fact that the earlier layers in a convolutional neural network detect the large scale features, and later layers look more at the details. For the purpose of our science cases, we are more interested in these larger scales.
\texttt{AstroMorphologyModel} is a simplified version of a truncated ResNet, which reduces training time. In our science cases, we found a similar performance when compared to an \texttt{NLayerResnet}.
When using the \byol and \texttt{BYOLTrainer} classes, the neural network can be freely defined and specified, whereas when using the top-level \texttt{astromorph} script, one can only choose from the two aforementioned networks (see Fig.~\ref{fig:astromorph-pipeline}).

\subsubsection{Learning rate and scheduler}
The learning rate governs how large the change is that is made to the model parameters at every optimization step, in the direction calculated by the optimizer.
This learning rate is set with the keyword \texttt{learning\_rate}, and should have a positive value.
If the learning rate is too low, it will take a long time to converge, but if the learning rate is too high, the model might never converge as the parameter values keep orbiting the minimum.
One way to circumvent this is by using a scheduled learning rate, where training starts with a relatively high learning rate to get the model parameters within range of the optimal values.
After a few epochs of this, the learning rate is decreased for finetuning of the parameters. For the training script, we provide an exponentially decaying learning rate, which can be enabled by setting \texttt{exponential\_lr = true} in the configuration file.
The decay parameter can be set with \texttt{gamma = <value>}, and should have a value between 0 and 1.

\subsubsection{Batch size}
The batch size parameter tells us how many samples of the training set are being processed between each optimization of the model parameters.
A batch size that is big will lead to needing many more epochs of training, as there are fewer optimization steps per epoch, and therefore the training process will take longer.
Too small of a batch size however will lead to erratic optimization, since a batch might not contain an adequate sampling of the training space, and each individual training step will overfit on a specific part of the training set.

\subsubsection{Epochs}
The number of epochs (complete passes through the entire training dataset), is controlled either via the epochs keyword in \texttt{example\char`_settings.toml} 
or by passing the epochs argument to \texttt{ByolTrainer}. We train for a fixed number of epochs and monitor performance to detect overfitting, for example when training error keeps decreasing while validation error starts to increase.

\subsubsection{Projection size and hidden size}
The embedding vector is reduced in size through a multilayer perceptron called the projector, denoted in Fig.~\ref{fig:astromorph-pipeline} with $g_{\theta}$. The projector is in essence a simple matrix multiplication, whereas the predictor is a fully connected multilayer perceptron with one hidden layer. Although both components are central to BYOL, their exact function is not fully understood; they are believed to help prevent representational collapse -- a failure mode in which the system converges to trivial constant outputs when trained only on positive pairs \citep[e.g., ][]{Richemond2023}.

The vector after projection is referred to as $z_\theta$. 
If this vector is too small, the model will be difficult to converge as it cannot contain all the representational information.
The projector also has a hidden layer, usually larger than the input layer. Expert users may tune the hidden and output sizes to explore representational capabilities; in the predictor, the output size is fixed to the projection dimension, but the hidden size remains adjustable.
    
\subsubsection{EWMA decay parameter}
Intuitively, the decay parameter $\lambda$ controls how rapidly the target network adapts to changes in the online network. Larger values of $\lambda$ result in a more slowly evolving target network, which stabilizes training, while smaller values allow the target network to track the online network more closely at the cost of reduced stability. Formally the decay parameter controls how the parameters of $f_\xi$ and $g_\xi$ are updated through an exponentially weighted moving average (EWMA) of the online component parameters (see Fig.~1).
If $f_{\theta, i}$ is the online neural network after optimization step $i$, the following equation applies for the target functions:

\begin{equation}
f_{\xi, i} = \lambda  f_{\xi, i-1} + (1 - \lambda) f_{\theta, i}
\end{equation}

Typically, we use a value of $\lambda = 0.99$, in accordance with \citep{grill2020}. This choice can be modified by advanced users, although the default is generally sufficient.

\subsubsection{Loss function}
The loss function used in BYOL is typically negative cosine similarity between the predictor and the target, which measures how well the prediction matches the target representation. This can also be modified by advanced users, though the default is usually sufficient.

\subsubsection{Optimizer}
The optimizing algorithm to find the best parameter values.
By default we use the ADAM optimizer, which is an adaptive stochastic gradient descent algorithm.

\subsection{Necessary housekeeping}
\subsubsection{Representation size}
The representation size is the number of dimensions in the embedding space, i.e. the length of the representation vector.
This number is fully determined by the architecture of the neural network.
However, we still need to explicitly pass this value to the BYOL framework.

\subsubsection{Stacksize}
Traditionally, optical astronomy uses single-channel images, whereas terrestrial images have three channels, one for each of the colours red, green, and blue.
Most pre-trained models are designed for and trained on these 3-channel RGB images and can therefore not ingest the single-channel images.
We work around this by stacking copies of the single channel to create an artificial 3-channel image.
This option can be enabled by using \texttt{stacksize = 3}. For data with a different number of channels, the user can instead define a neural network architecture with a corresponding number of input channels.

\subsubsection{Datasets}
PyTorch provides the Dataset primitive to easily store data and their attributes.
We have defined a FitsFilelistDataset class that does this for FITS files.
Initialization is done with a simple list of file locations, and the I/O is handled by the class itself.
Also stacking the images according to the \texttt{stacksize} parameter, and data augmentation by mirroring and 180-degree rotation is handled by this class.

\begin{figure*}
\centering
\includegraphics[width=1.0\hsize]{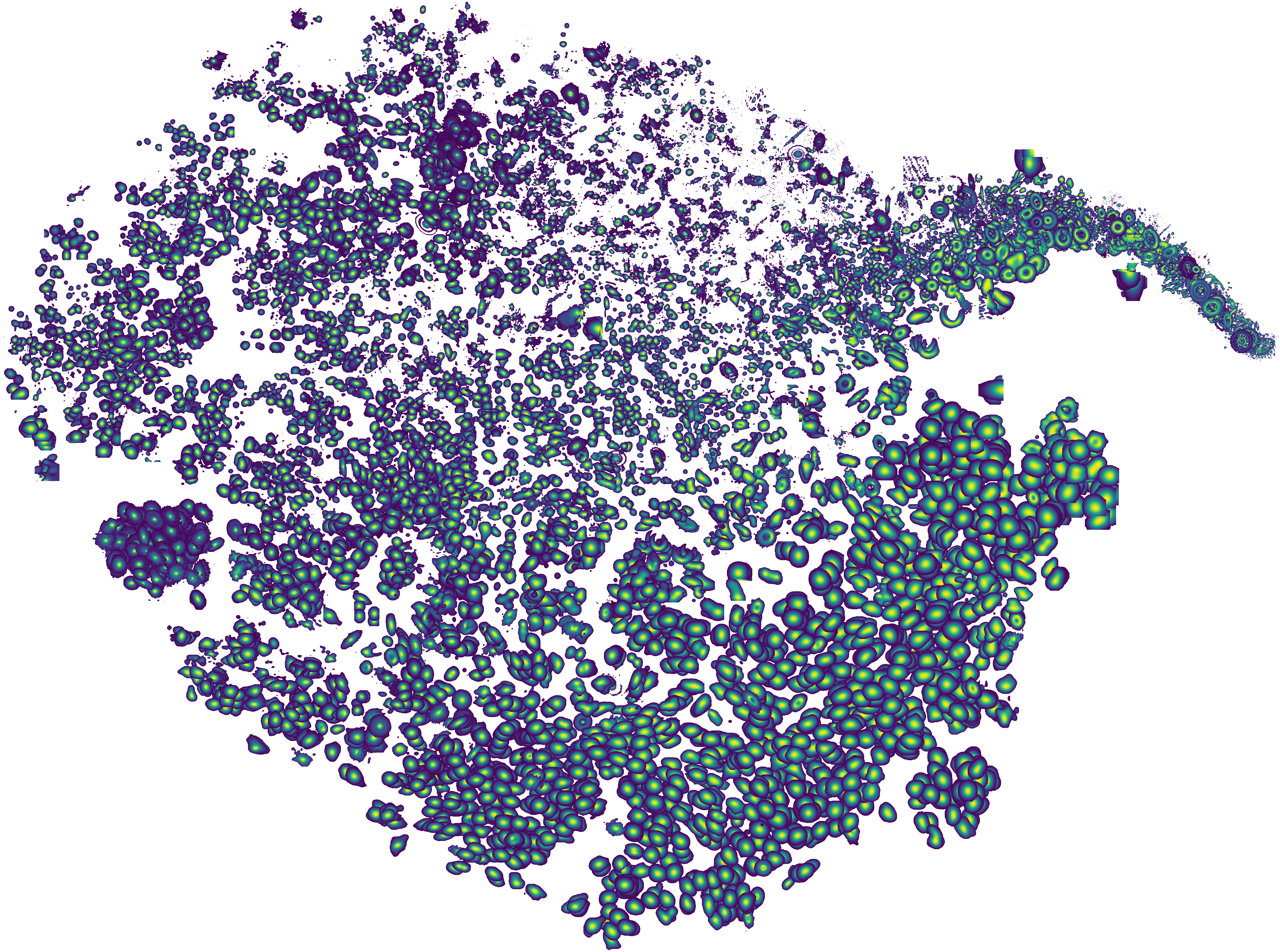}
\caption{t-SNE projection of embedding vectors for ALMA continuum images selected from the ASA. Each point is represented by a thumbnail of the corresponding FITS image. For clarity, low-emission regions have been masked using an adaptive threshold derived from a Gaussian fit to the core of the distribution of normalized pixel values. Sources with stronger emission appear on top of weaker ones in the plot to enhance visibility.} 
\label{fig:proj_ALMA}
\end{figure*}

\section{Science demonstration}
\label{sec:results}
Below, we demonstrate the application of the standard \texttt{astromorph} pipeline to two contrasting science cases, using the out-of-the-box configuration in each instance. Although both demonstrations are based on 2D image data, they represent fundamentally different usage modes: the analysis of heterogeneous archival data and the analysis of a more homogeneous survey context. The goal here is not to provide a comprehensive scientific analysis, but rather to illustrate the flexibility and scope of the method. More detailed investigations of both science cases will be addressed in future work.

In the first case, we focus on sub-mm continuum images of low-mass star forming regions and protoplanetary disks observed with ALMA. Since ALMA observations rarely cover a large fraction of the sky and can cover a wide range of angular resolutions, it is reasonable to assume that an automated organization based on the morphology of these images does not provide an unbiased view of the true populations statistics or distribution of object types. Instead, unsupervised exploration in this context is most valuable for identifying morphological diversity, grouping of similar objects/observations, and enabling efficient searches for targets with particular structures of interest. 

In contrast, our second science case involves the analysis of infrared dark clouds observed with Spitzer, where individual objects are extracted from a single, contiguous dataset covering a larger region of the sky. Consequently, an automated morphological organization is expected to offer valuable insights into the true distribution of various objects and underlying physical processes governing them. 

For both cases, we use the full convolutional backbone of the ResNet18 architecture (layers 1 through 4). For the BYOL settings, we use a representation of 512 and a projection size of 128. The network is trained for 15 epochs, a value chosen empirically to provide a good trade-off between representation quality and the risk of overfitting. 

\subsection{Disks}
\label{sec:disks}
While organization of ALMA data may not yield precise information regarding the true distribution of various objects, it can nevertheless serve as a valuable tool for obtaining a general overview of the types and frequencies of objects targeted throughout the operational period of ALMA. 

A key difference between this demonstration and that presented in Sec.~4.2 lies in the nature of the ALMA archival data, which exhibits substantial variability in image quality (manifested as differences in noise levels, imaging artifacts, and field-of-view, etc.) between different observations. For that reason, the data needs to be curated. This curation includes level correction to avoid variations in noise levels in primary beam corrected images, noise clipping at the 3-sigma level, DBSCAN to identify and extract sources, and finally, the generation of FITS files with 100 x 100 pixels. All continuum images available in the ALMA Science Archive (ASA) as of January 20, 2025, containing the keywords "winds, jets and outflows," and "disks around low-mass stars," were downloaded and curated. These 5941 images were then processed through \texttt{astromorph} for classification. It is worth to mention, that also larger images were tested when carrying out the classification (200 x 200 pixels), but that makes no significant difference to the outcome presented here.

\begin{figure*}[h]
\centering
\includegraphics[width=1.0\hsize]{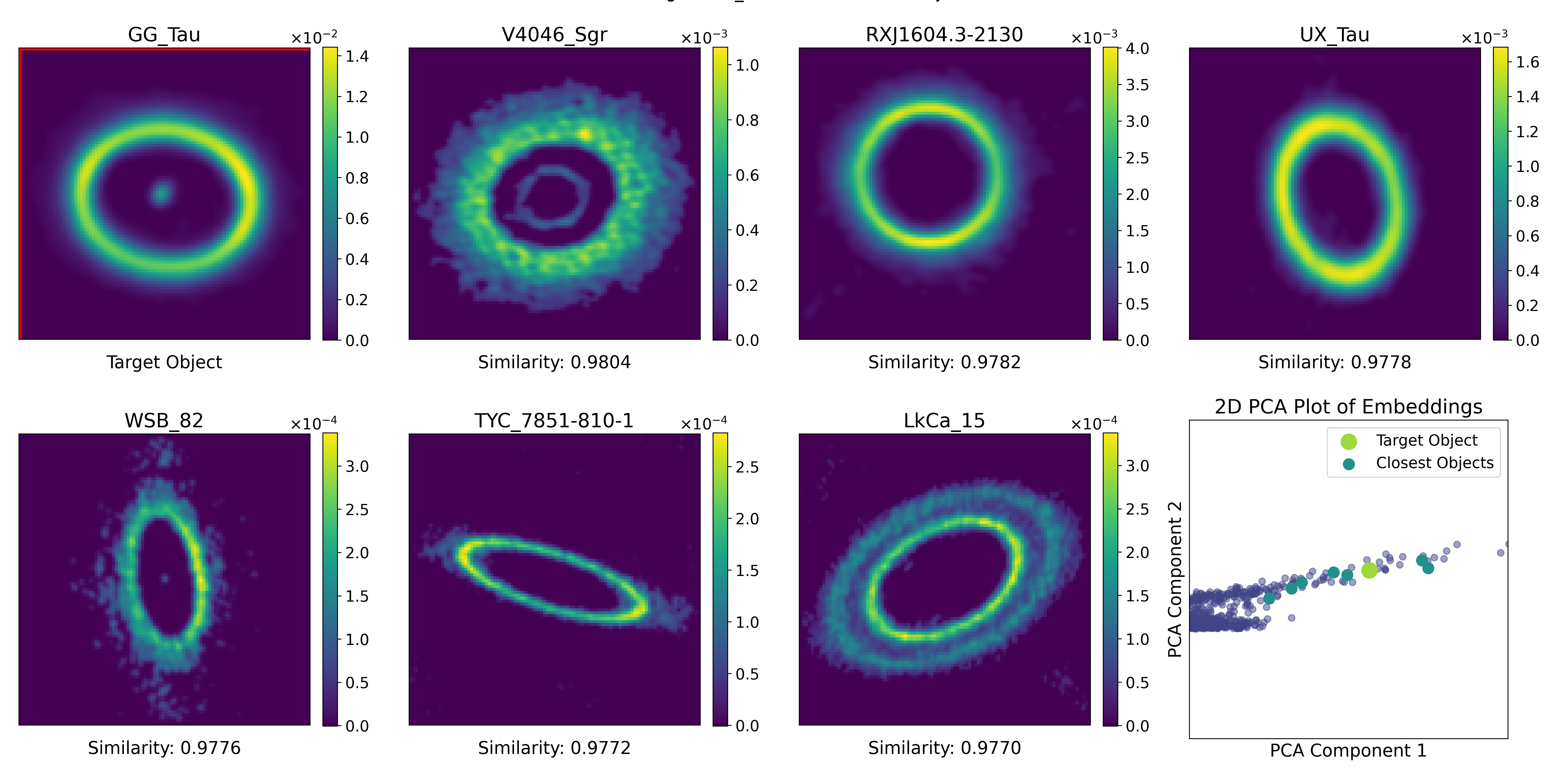}
\caption{ALMA continuum image of GG Tau (upper left) and the six most morphologically similar images, identified using cosine similarity. Each thumbnail shows the flux density in Jy beam$^{-1}$. Lower right panel presents a 2D PCA projection of embedding vectors, with the selected object and its closest matches highlighted.} 
\label{fig:similarityfigure}
\end{figure*}
To visualize the distribution of all images in the dataset, embedding vectors are extracted and projected into two dimensions using t-Distributed Stochastic Neighbor Embedding (t-SNE) \citep{vanDerMaaten2008}.
The result of this dimensionality reduction is presented in Fig.~\ref{fig:proj_ALMA}, where the noise has been removed from each image by applying an adaptive thresholding procedure: after normalizing the pixel values, a Gaussian distribution is fitted to the core of the intensity histogram -- defined as values within a narrow range around the mode -- and all pixels below a threshold set at 5$\sigma$ above the fitted mean are masked out. For a 2D presentation, t-SNE is advantageous due to its ability to preserve local structures and reveal clusters in high-dimensional data.

Inspection of Fig.~\ref{fig:proj_ALMA}, shows that objects fall into a roughly heart-shaped manifold with multiple lobes, suggesting the presence of distinct categories. Meanwhile, the lack of sharp boundaries and the gradual transition between different morphologies may reflect that data curation was largely consistent and reasonable. If we take a closer look at the distribution, it is clear that a large proportion of the continuum data observed with ALMA is unresolved. On the other hand, there are also parts of the dataset that are resolved and where similar morphologies are grouped together. One such example is ring-like structures in protoplanetary disks, which to a reasonable extent occupy nearby positions in t-SNE space. 

A natural question arising from this is whether \texttt{astromorph} can be utilized to analyse one of these subcategories in greater detail. Specifically, we deem it valuable to assess whether a given observation can be used to identify other observations with similar morphological characteristics. In fact, \texttt{astromorph} allow us to address this question in a similar manner as the organization itself. Whether we want to find similar objects or organize all objects, we base our approach on the list of embeddings produced by the pipeline. As a test case we selected the circumbinary structure in GG Tau \citep{Dutrey1994}, which is scientifically interesting because it offers a rare glimpse into how planets might form in complex, dynamic environments influenced by multiple stars. Based on a selected continuum image of GG~Tau, we compute similarity between objects from the embedding vectors using either cosine similarity or normalized euclidean distance. A search is made for objects matching a given particular source name and retrieves their embedding vectors. The most similar images are then identified based on the chosen distance metric. In Fig. \ref{fig:similarityfigure} we provide an example showing the most similar images to one selected continuum image of GG Tau from dataset \texttt{uid\_\_\_A001\_X133d\_X1f29}, filename \texttt{member.uid\_\_\_A001\_X133d\_X1f29.GG\_Tau\_sci.spw25\_27\_29\_31\_33\_35\_37\_39\_41\_43.cont.I.pbcor.fits}, retrieved from ALMA project 2018.1.00532.S (all other continuum images of GG~Tau are excluded), which exhibits ring like structures. The tests show that it is indeed possible to find objects that exhibit similar morphological structures. Meanwhile, it is worth mentioning that these findings do not in any way prove that the underlying physical processes in those similar images are the same, and found images should therefore be seen as a starting point for further studies. As it is  obvious from Fig.~\ref{fig:similarityfigure}, found ring-like structures are in some cases very different morphologically from each other, and given their diversity, it is not entirely trivial for \texttt{astromorph} -- with this dataset at hand -- to distinguish, for example, whether there is one or multiple rings, or if the rings have different morphological appearances.

\subsection{Morphology of molecular clouds in the Milky Way}
\label{sec:clouds}
\begin{figure*}
\centering
\includegraphics[width=1.0\hsize]{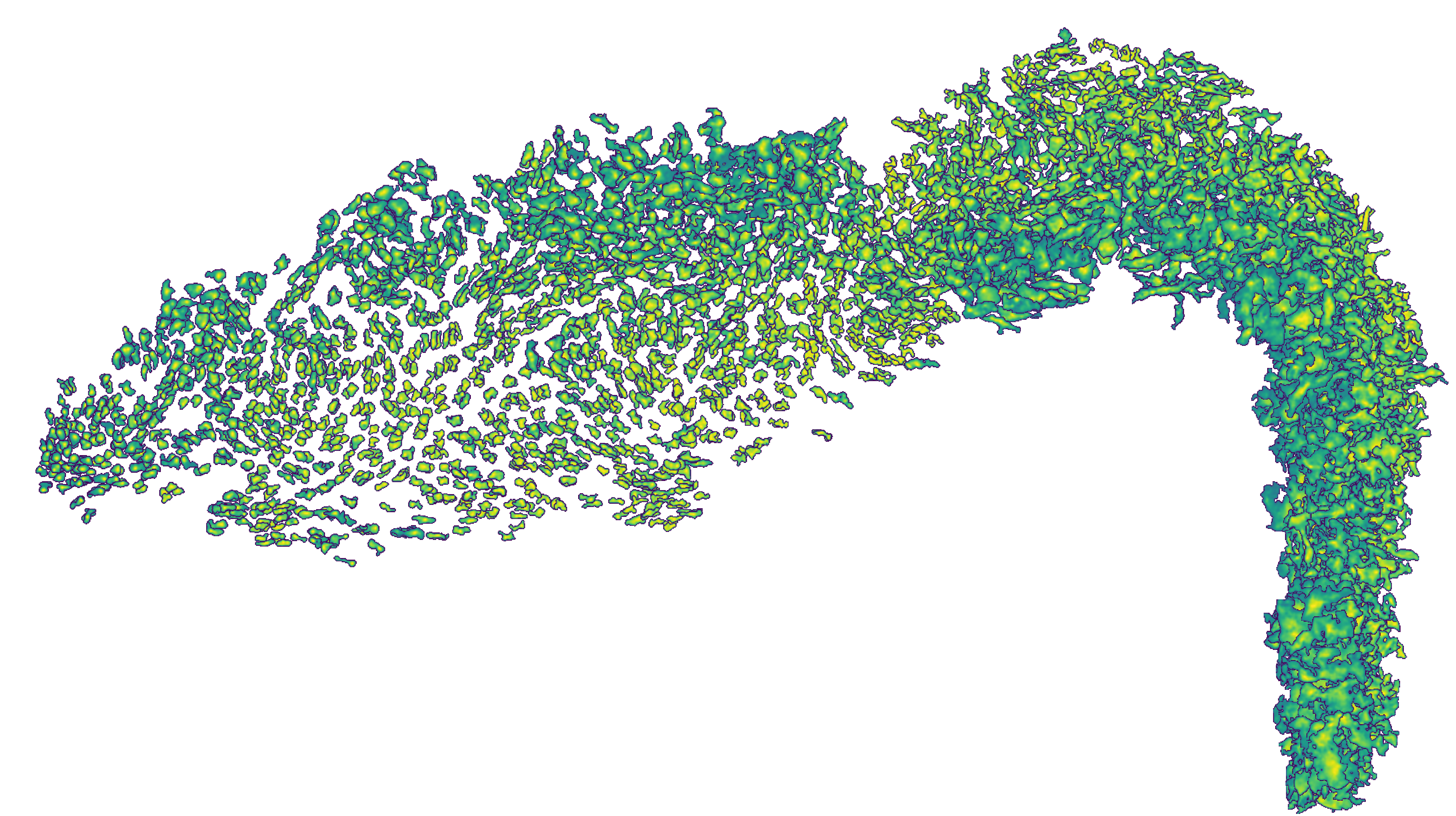}
\caption{Same as Fig.~\ref{fig:proj_ALMA} but for cloud data obtained with Spitzer. }
\label{fig:proj_CLOUD}
\end{figure*}

We demonstrate the utility of \texttt{astromorph} for analysing morphology of molecular clouds. We
know today that new stars form in the cold and dense gas clouds of the interstellar medium. One fundamental, yet open
question in this topic is understanding the life cycle of individual clouds during their evolution from
quiescence to star formation. This question translates to trying to understand the roles of different
forces/phenomena, such as gravity, turbulence, and thermal pressure, during the cloud evolution.

One avenue to address the question is to look for fingerprints of those processes in the cloud structure and morphology. It is generally considered that the different processes affect the morphological properties of gas and that observations of those properties can constrain the underlying processes. Such measures include, for example, central concentration, elongation, orientation, area-perimeter ratio, central moments, fragmentation statistics, one- and two-point statistics of the structure, power spectra, and many others \citep[e.g.,][]{Larson1981, Kainulainen2009probing, Schneider2011, Koch2019, Hacar2023Review}.   

In this demonstration, we address a question: Will the BYOL method and the \texttt{astromorph} package produce scientifically interesting/relevant embeddings for the analysis of the molecular cloud morphology? This question is inherently exploratory and it is not a priori clear what all ‘scientifically interesting’ may include; it remains to be decided subjectively by the user. Generally, interesting embeddings should contain information that enables gaining insight into the physical properties of the clouds via their morphology.   

We use as data a sample of molecular clouds identified from the Milky Way by the PROMISE survey (Kainulainen et al. in prep.). In short, the data comprise column density maps of thousands of molecular clouds originally identified as infrared dark clouds in the Spitzer/GLIMPSE survey data at 8 $\mu$m wavelength. The column densities have been derived using a combination of the 8 $\mu$m shadowing signature and dust emission data from the Herschel satellite. The details of the  data set are not important for the goals of this demonstration; the survey and the data set will be described and made publicly available via Kainulainen et al. (in prep.). For this demonstration, we merely consider the data as a set of morphologically complex images. 

The input sample for \texttt{astromorph} consists of a collection of about 2\,900 molecular cloud region maps. These maps vary around some tens and two hundred pixels in size and vary greatly in their morphology (see Fig.~\ref{fig:proj_CLOUD}). The data set was preprocessed so that we imposed a low threshold cut to isolate continuous structures from the maps, and we logarithmically scaled the data to emphasize the overall structure of the clouds as opposed to being dominated by the local maxima.   

We analysed the data set using the \texttt{astromorph} pipeline (\ref{sec:astromorph_pipeline}) with the exact same settings as in the disk case. 
Figure \ref{fig:proj_CLOUD} shows the t-SNE projection of the embeddings. 
\begin{figure*}
    \centering
    \includegraphics[width=0.99\textwidth]{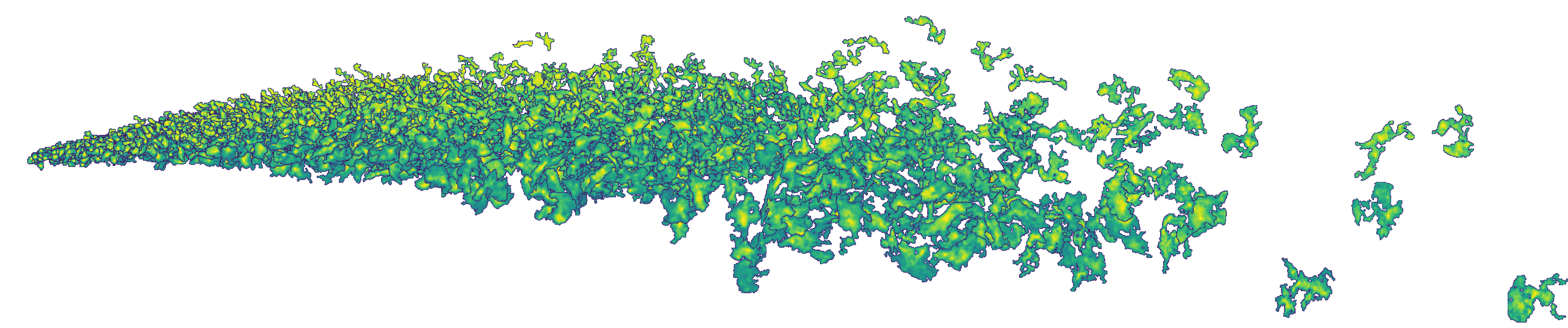}%
    \\
    \vspace{2cm}
    \includegraphics[width=0.99\textwidth]{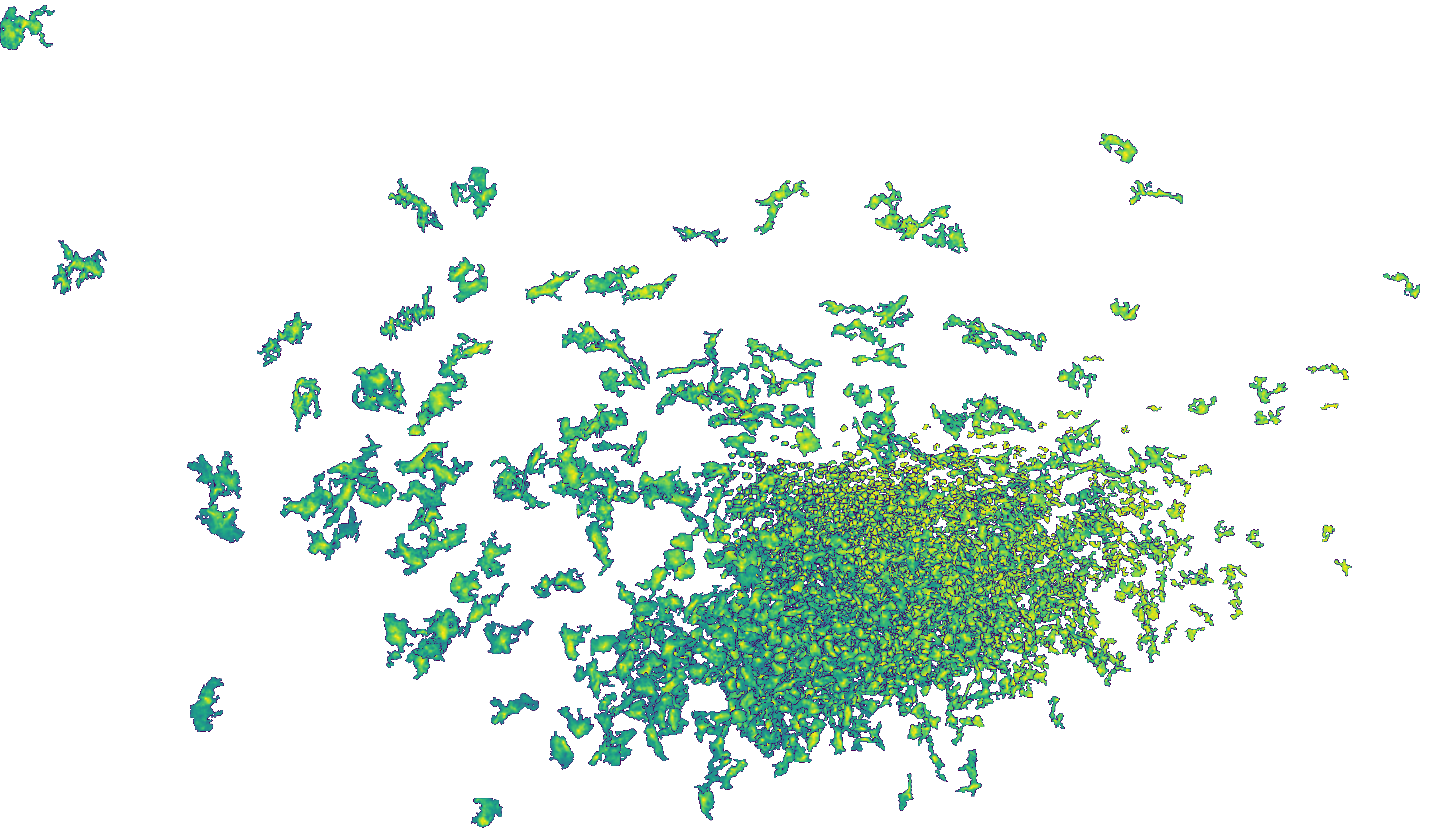}%
    \caption{Principal component analysis of cloud morphologies. \textit{Upper panel: }PC1 vs. PC2, where PC1 primarily seems to trace cloud size and PC2 reflects the distribution of the emission. \textit{Lower panel: } PC2 vs. PC3 where PC2 again represent emission distribution, while PC3 seems to capture some measure of morphological complexity. }
    \label{fig:PCA_clouds}
\end{figure*}

The t-SNE representation gives rise to some immediately interesting analysis avenues. First, in comparison to the representation of the disk data (Fig. \ref{fig:proj_ALMA}), the embeddings are distributed in a more continuous manner, along a curved manifold. Along this manifold, several morphological features can be seen; for example, the filling factor of a structure (size), and elongation clearly play a role. We note that, while rotation-based augmentation is intended to reduce sensitivity to orientation, residual orientation-related effects may still be discernible, potentially due to correlations with other morphological features and the difficulty of enforcing strict rotation invariance for structurally complex data.

The visual appearance of the t-SNE representation is somewhat different than in case of the disk data in which the projection is more segregated into "islands" of different morphological features. One avenue to explore and exploit this difference could be a physical interpretation that the cloud data is better described with continuous (scaling) relations and sequences, rather than discrete classes as in the case of the disk data. Such direction would immediately link the information from embeddings to a physically interesting  interpretations.

Another example of possible analysis directions is that the density of objects in the projection reflects how common the represented morphologies are. For example, the small, isolated elongated clouds are clearly rarer than clouds with complex morphology. Of course, an analysis going to this direction should address observational biases and projection effects affecting the cloud morphology, but in general, the projection of the embeddings clearly facilitates such approach.    

The t-SNE representation of the cloud data is perhaps less straightforward to interpret than that of the disk data. We therefore continue the exploration with another analysis and illustration, to further demonstrate scientific potential of the approach. For this, we performed a principal component analysis (PCA) of the embeddings. The PCA represents multidimensional data (here, the embeddings) with 
a set of orthogonal axes that capture the dominant sources of variation. Fig.~\ref{fig:PCA_clouds} shows the result of this analysis. Qualitatively, the first principal component (PC1) seems to primarily trace the overall size of the clouds, while the second component (PC2) reflects the distribution of the emission, separating more uniformly filled clouds from those that are more peaked or centrally concentrated. The third component (PC3) seems to capture some aspect of the morphological complexity, distinguishing smoother and rounder structures from more filamentary or branched ones. Taken together, the PCA results show that the variation in cloud morphologies is complex, but the first components already appear to reflect meaningful physical properties. This initial indication suggests that there is underlying physical information encoded in the embeddings, and that a proper, thorough analysis (of also higher-order components and their correlations with physical parameters) will be a fruitful avenue. For instance, it would seem to enable quantification of the importance of morphological parameters, which in conjunction with auxiliary data (e.g., star formation activity) would link the cloud morphology to evolution of the star-forming Inter Stellar Medium (ISM).

\subsection{Outlook and possible future science applications}
The versatility of \texttt{astromorph} opens up a wide range of promising directions beyond the examples presented here. A natural next step would be a deeper exploration of the ASA itself. As more continuum observations become public, the archive increasingly contains underexplored or rare categories of sources. With relative little manual input, \texttt{astromorph} could be used to scan for outliers or examples of rare morphologies. A similar promising direction lies in deeper exploration of the Spitzer and Herschel molecular cloud data with the new perspective enabled by a ``blind'' approach to molecular cloud morphology, as opposed to identifying or extracting pre-defined shapes such as cores, filaments, etc.

Beyond the above examples, new opportunities are emerging with JWST. For instance, MIRI and NIRCAM have produced highly structured views of young protostellar regions \citep[e.g.][]{2024ApJ...967..110L} and their surrounding environments \citep{2022ApJ...936L..14P}. Such images often reveal complex combinations of scattered light, outflows and dust extinction, resulting in diverse morphologies that are difficult to organize in a systematic way. Applying \texttt{astromorph} here, could reveal trends within this diversity, grouping together similar YSO's, component shapes and trends therein, and ultimately linking morphological traits to physical properties. In the area of extragalactic ISM and star formation, JWST-PHANGS survey \citet{Lee2023, Williams2024} and the large number of linked PHANGS surveys\footnote{see~\url{https://www.phangs.org/projects}. } are naturally well-suited large data sets to apply \texttt{astromorph} to study ISM morphology and its relation to star formation and galaxy evolution.

Another avenue lies in large-scale survey data both from former and future facilities. Future SKA maps will trace the atomic and molecular gas in in the Milky Way in great detail. By studying such plethora of complex structures with \texttt{astromorph}, it may be possible to trace how the morphology of gas structures varies across different early stages of ISM evolution, in different galactic environments, or to classify the morphology of bubbles, shells or shock fronts driven by stars or supernovae. 

It is also worth emphasizing that, although we currently apply \texttt{astromorph} only to 2D data, there is no limitation preventing the extension of the use to higher-dimensional inputs. In particular, the vast repositories of spectral line data cubes represent an extraordinary opportunity. By incorporating spectrally and spatially resolved datasets one could capture not only the projected morphology of objects, but also its kinematic and/or chemical structures. This opens the door to a far richer exploration of morphological diversity, where for example dynamic state may be encoded directly into the learned representations. 

Moreover, the framework is not limited to astronomical applications. For instance, \texttt{astromorph} has already been used at Chalmers University of Technology to analyse ship movements in the Baltic Sea, where vessel tracks and coastlines formed two different input channels, thus making use of the 3D capability of the package. 

In all of these cases, we envision \texttt{astromorph} being a flexible tool that can be adapted across wavelength and science domains, and even beyond astronomy, offering a new way to navigate the increasingly image-rich future.

\section{Conclusion}
\label{sec:conclusions}
We have presented \texttt{astromorph}, a Python package designed to apply the BYOL method for self-supervised learning to astronomical data. The package enables users to extract meaningful morphological embeddings from datasets without requiring labelled examples. 

The package is tailored specifically for astronomical use, and we provide three options for running it, depending on the expertise level of the user. This setup was developed to help users overcome the often-steep learning curve associated with the use of machine learning frameworks.

We have demonstrated the application of \texttt{astromorph} in two science cases. In ALMA continuum observations of low-mass star-forming regions, we showed how the approach identified morphological similarities across datasets -- such as ring-like structures in disks -- despite variability in noise levels and resolution. When applied to Spitzer- and Herschel-based molecular cloud maps, the approach captured structural differences between clouds, demonstrating promising avenues to explore large-scale morphological patterns and their connections to underlying physical processes like fragmentation, gravity and turbulence, and star formation.

\begin{acknowledgement}
MCT acknowledges support of Onsala Space Observatory national infrastructure. Onsala Space Observatory is supported through Swedish Research Council grant No 2019-00208. The project was enabled by resources provided by Chalmers e-Commons at Chalmers. We acknowledge the use of ChatGPT (OpenAI, versions 4 -- 5) for language editing and clarity suggestions in the drafting of this manuscript. \textit{Author contributions: }All authors contributed to the development and writing of this paper. The authorship list is given in two alphabetical groups. The first group (PB, JK, MCT) contributed equally: they initiated the project, carried it through, wrote the manuscript, performed the scientific analysis, developed code for the curation of input data, and contributed to the development of the \texttt{astromorph} package. The second group (LB and OML) was responsible for the development of the \texttt{astromorph} package, and contributed to the manuscript. 
\end{acknowledgement}

\bibliographystyle{aa}
\bibliography{references}

\label{LastPage}
\end{document}